\documentclass[aps,prl,nofootinbib,twocolumn,floatfix,
letterpaper,superscriptaddress,showpacs]{revtex4}
\usepackage{graphicx}\usepackage{natbib}\usepackage{amsmath}
\begin{document}

\title{New Constraints on the Highest-Energy Cosmic-Ray Electrons and Positrons}

\author{Matthew D. Kistler}
\affiliation{Center for Cosmology and Astro-Particle Physics and Department of Physics, Ohio State University, Columbus, Ohio 43210}

\author{Hasan Y{\"u}ksel}
\affiliation{Bartol$\,$Research$\,$Institute$\,$and$\,$Department$\,$of$\,$Physics$\,$and$\,$Astronomy, University$\,$of$\,$Delaware,$\,$Newark,$\,$Delaware$\,$19716}

\date{December 1, 2009}

\begin{abstract}
At energies above a few TeV, no measurements of the cosmic-ray electron spectrum exist yet.  By considering the similarity of air showers induced by electrons and gamma rays as seen by ground-based arrays, we use published limits on isotropic gamma-ray fluxes to place first constraints on the $> 10$~TeV electron spectrum.  We demonstrate that, due the proximity of known sources, the flux of such electrons (and positrons) can be large.  We show how these smoothly connect to lower-energy positrons measured by PAMELA and relate to exciting new indications from {\it Fermi}.
\end{abstract}

\maketitle

\textit{Introduction.}---
Interest in the cosmic-ray electron spectrum at Earth is at an all-time high, arising from both astrophysical and dark matter-related concerns~\cite{Adriani:2008zr,Fermi:2009zk,Aharonian:2008aaa,Aharonian:2009ah,Yuksel:2008rf,Hooper:2008kg,Profumo:2008ms,Blasi:2009bd,Chang:2008zzr,ArkaniHamed:2008qn}.  In the GeV energy range, much greater clarity than what had existed in preceding decades has been brought by the PAMELA~\cite{Adriani:2008zr} and {\it Fermi}~\cite{Fermi:2009zk} space missions.

Such measurements, which observe electrons and positrons directly, become difficult above $\sim 1$~TeV due to their fixed detector areas relative to declining particle fluxes.  Using the indirect technique of observing atmospheric air showers, HESS has pushed the energy frontier up to several TeV~\cite{Aharonian:2008aaa}, although above this, no electron measurements have been reported.  With the historical lack of local multi-TeV gamma-ray sources and the soft spectra of secondary $e^\pm$ resulting from $p$--$p$ scattering~\cite{Moskalenko:1997gh}, little if any signal may have been expected.

Due to the similarity of the electromagnetic showers produced in the atmosphere by energetic electrons and gamma rays, they are nearly inseparable via ground-based observations~\cite{Hofmann:2006wf,Aharonian:2001ft}.  Here, we translate published limits on isotropic gamma-ray fluxes (e.g.,~\cite{Aharonian:2001ft,Chantell:1997gs,Schatz:2003aw,Gupta}) to constrain the $e^- + e^+$ spectrum to $>\,$PeV energies.

We further show that our derived limits are relevant in light of recent measurements by {\it Fermi} and TeV gamma-ray telescopes, particularly the discovery and detailed observations of high-energy sources both near and far that indicate $e^\pm$ production in the GeV and TeV regimes.  Considering these observations, with an improved analytical treatment of $e^\pm$ propagation to calculate expected source contributions at Earth, we conclude that it would be surprising if the influence of a nearby pulsar was {\it not} present in cosmic-ray positron data.

\begin{figure}[b!]
\includegraphics[width=\columnwidth,clip=true]{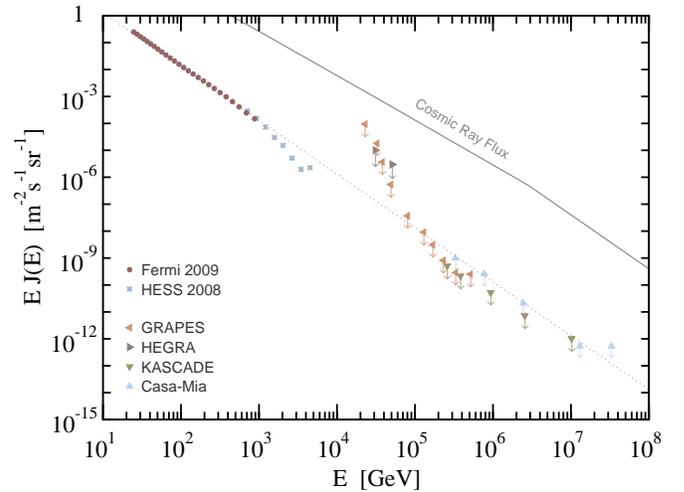}
\caption{The cosmic-ray electron spectrum at Earth.  Shown are direct $e^- + e^+$ measurements from {\it Fermi}~\cite{Fermi:2009zk}, measurements of $e^- + e^+$ based on showers from HESS~\cite{Aharonian:2008aaa,Aharonian:2009ah}, and a baseline $\propto E^{-3}$.  These can be compared to our limits derived from gamma-ray experiments~\cite{Aharonian:2001ft,Chantell:1997gs,Schatz:2003aw,Gupta} at $>10$~TeV.
\label{limits}}
\end{figure}

\textit{New Cosmic-Ray $e^\pm$ Limits.}---
While the $\sim 1$~m$^2$ Large Area Telescope (LAT) of {\it Fermi}~\cite{Fermi:2009zk} has brought a sharper picture of the $e^- + e^+$ spectrum up to $1$~TeV, to progress further requires a much larger effective area for particle collection.  This can be accomplished by, in lieu of direct particle identification, examining the showers that result when energetic particles scatter in the upper atmosphere.

To observe an electromagnetic shower (initiated by a gamma ray or electron) requires rejecting the large background due to cosmic-ray protons.  Ground-based air Cherenkov telescopes (ACTs) operate by imaging the shower and observing the differences between hadronic and electromagnetic cascade development~\cite{Aharonian:2008zz}, which allowed HESS to measure $e^- + e^+$ up to several TeV (Fig.~\ref{limits}).  Alternatively, detectors that operate by directly detecting the long-lived products of air showers as they reach the ground make use of the low muon content of electromagnetic showers relative to hadronic events~\cite{Aharonian:2008zz}.

Searches for isotropic fluxes of gamma rays have been conducted over a wide range of energies by examining muon-poor showers~\cite{Chantell:1997gs,Schatz:2003aw,Gupta}.  While these have typically only resulted in upper limits, they can be valuable.  Since a gamma ray first produces an initial $e^\pm$ pair to begin a cascade, its shower will look very similar to that of an electron of equivalent energy, only located $\sim \,$one radiation length deeper in the atmosphere~\cite{Hofmann:2006wf}.  HESS electron data can thus be regarded as upper limits on an isotropic TeV gamma-ray flux~\cite{Aharonian:2008aaa,Kistler:2009xf}, while air shower arrays do not have hope of exploiting this subtle difference, rendering diffuse gamma rays and electrons inseparable~\cite{Aharonian:2001ft}.

By using isotropic gamma-ray limits from air shower arrays, we derive new limits on the cosmic-ray electron spectrum in a regime currently lacking constraints.  These limits are often quoted as the fraction of measured gamma-ray to proton intensity, $I_\gamma/I_{\rm CR}$.  We use the cosmic-ray nuclei spectrum~\cite{Hillas:2006ms} in Fig.~\ref{limits}.  In the range $10-10^5$~TeV, we obtain electron limits from HEGRA~\cite{Aharonian:2001ft}, CASA-MIA~\cite{Chantell:1997gs}, GRAPES~\cite{Gupta}, and KASCADE~\cite{Schatz:2003aw} data, leading to the limits in Fig.~\ref{limits}.

We have conservatively assumed that the electron fraction of the electromagnetic showers seen is $100\%$.  It is likely that dedicated re-analyses of the data can strengthen these constraints, and an analysis of Milagro~\cite{Abdo:2009ku} data could probe the 10--100~TeV region.  Even so, we see that our limits are already competitive when compared to previous measurements at lower energies.

\textit{Nearby $e^\pm$ Factories.}---
Due to their short lifetime against radiative losses, any electron flux measured at $E>10$~TeV must be produced in the very-recent history of a nearby source.  For this reason, we will consider sources within $\sim 500$~pc that still exhibit evidence of particle acceleration to very-high energies, using more distant sources for added guidance.  Observations, principally of extended gamma-ray emission, have recently revealed good reasons to believe that $e^\pm$ are being produced, accelerated up to multi-TeV energies, and escaping from the high-energy-density environments of pulsars:

$\bullet$ HESS has intensively observed the distant ($\sim\,$4~kpc) pulsar wind nebula (PWN) HESS~J1825--137~\cite{Aharonian:2006zb}.  This included untypically-long exposure times due to the PWN being serendipitously within the field-of-view of the variable microquasar LS~5039, for which repeated measurements of the light curve were taken.  These revealed an extended wind of $\gtrsim\,$10~TeV $e^\pm$ containing $\gtrsim\,$$10^{48}$~erg reaching $\gtrsim 100$~pc in only $\sim\,$20,000~yr~\cite{Aharonian:2006zb}.  This indication of a pulsar {\it ``TeV mode''} carrying $\gtrsim\,$$10^{48}$~erg of $e^\pm$ to great distances likely would have been missed without these deep observations, and thus may be common.

$\bullet$ The proximity (290~pc) and relative youth ($\sim\,$11,000~yr) of the Vela pulsar has permitted detailed observations of its PWN, Vela~X.  There, HESS inferred a population of $e^\pm$ with energies reaching $\sim 100$~TeV~\cite{Aharonian:2006xx}.  Modeling source $e^\pm$ injection spectra as $d\dot{N}/dE \propto E^{-\alpha} e^{-E/E_{\rm max}}$, the HESS measurements imply $\alpha=2$ and $E_{\rm max}=70\,$~TeV for Vela~X~\cite{Aharonian:2006xx}.  Intriguing evidence has also been obtained at lower energies, with a distinct population of multi-GeV, radio-synchrotron emitting $e^\pm$ (with an $E^{-1.8}$ spectrum) modeled in Ref.~\cite{deJager:2008ni}.  The (preliminary) {\it Fermi} discovery of extended GeV gamma rays in Vela~X confirms this and, importantly, indicates a high-energy cutoff in the spectrum at $\sim \,$130~GeV~\cite{Lemoine}.  The amplitude of this signal implies that a pulsar {\it ``GeV mode''} can also produce $\gtrsim\,$$10^{48}$~erg of $e^\pm$.  This suggests that the total composition of Vela~X includes $\sim \,$100~GeV $e^\pm$ accelerated by the pulsar itself --- likely associated with pulsed GeV gamma rays --- and TeV $e^\pm$ from shock acceleration of the pulsar wind.  As observed, the GeV component contains $\sim \,$100 times more energy, although multi-TeV particles may have already exited the system.

$\bullet$ The discovery of extended $\sim\,$35~TeV gamma-ray emission by Milagro~\cite{Abdo:2009ku} surrounding the nearby ($\sim\,$200~pc) pulsar Geminga indicates a close, active source of $e^\pm$~\cite{Yuksel:2008rf}.  These data imply a firm lower limit on the maximum particle energy of $\gtrsim 100$~TeV, and may approach $1000$~TeV.  In all of these systems, the implied particle multiplicities needed to account for the gamma rays via inverse-Compton scattering require $e^\pm$ pair production~\cite{Yuksel:2008rf,deJager(2007)}.  Since, like Vela, Geminga possesses bright pulsed GeV emission, it may as well have resulted in an abundance of $\sim \,$100~GeV $e^\pm$.  While Vela is too young for GeV $e^\pm$ to have reached us yet (as we will soon see), the greater age of Geminga ($\sim\,$300,000~yr) may allow for a direct test of the commonality of dual high/low-energy pulsar $e^\pm$ populations.  To determine this, we first address the propagation of highly-energetic $e^\pm$ in the Galaxy.

\textit{TeV $e^\pm$ Propagation.}---
In spherically symmetric geometry, the diffusion equation governing the particle density at a given location/time/energy, $n(r,t,E)$, is~\cite{Ginzburg(1964),Strong:2007nh}
\begin{equation}
  \frac{\partial n}{\partial t}= \frac{{D}(E)}{r^2} \frac{\partial}{\partial r}r^2
       \frac{\partial n}{\partial r} + \frac{\partial}{\partial E} [b(E) \, n] + Q\,,
\label{eq:diff}
\end{equation}
with energy losses parametrized as $b(E)=-dE/dt$, diffusion coefficient $D(E)$, and source term $Q$.  We first consider a single burst from a point source with a generated particle spectrum $dN/dE_g$, so that $Q(r,t,E_g) =  \delta(r) \, \delta(t) \, dN/dE_g$.  Using the Syrovatskii propagator~\cite{Ginzburg(1964)} (as in \cite{Aharonian(1995),Yuksel:2008rf}) yields
\begin{equation}
  n_S(r,t,E)= \frac{e^{-r^2/r_{dif}^2}} {\pi^{3/2} \, r_{dif}^3} \frac{dN}{dE_g}\frac{dE_g}{dE}
\,,
\label{eq:syrovatsky}
\end{equation}
where ${dE_g}/{dE}$ maps the energy at generation $E_g(E,t)$ to the observed $E$ after losses.  In the high-energy regime of interest here, most relevant are inverse-Compton losses on the CMB (energy density $\sim 0.3$~eV$\,$cm$^{-3}$) and synchrotron losses due to a $\sim 5\, \mu$G magnetic field ($\sim 0.2$~eV~cm$^{-3}$), so that $b(E)=b_0\,E^2$ with $b_0\simeq 5 \times 10^{-16}$~s$^{-1}$~GeV$^{-1}$.  Integration yields $1/E=1/E_g+b_0 t$, so that $dE_g/dE=(E_g/E)^2$.

The diffusion radius is $r_{dif}(E,t)=2\sqrt{\lambda(E,t)}$, with
\begin{equation}
  \lambda(E,t) = \int_0^t dt' D[E(t')] = \int_E^{E_g} d E' \frac{D(E')}{b(E')}\,.
\label{eq:lambda}
\end{equation}
We parametrize the diffusion coefficient as $D(E)= d_0 (1+E/E_0)^\delta$, with $E_0\simeq 3$~GeV, and consider values of $d_0 = 2 \times 10^{28}$, $4 \times 10^{27}\,$cm$^2$s$^{-1}$, $\delta =\,$0.4, 0.5~\cite{Strong:2007nh,DiBernardo:2009ku} corresponding to ``fast'' and ``slow'' models.  Since $D(E) \simeq d_0 (E/E_0)^\delta$ at energies $E>E_0$, the diffusion radius simplifies to
\begin{equation}
  r_{dif}(E,E_g) \simeq 2\left( \frac{d_0}{b_2}
      \frac{E^{\delta-1}-E_g^{\delta-1}}{(\delta-1)E_0^{\delta}}\right)^{1/2} \,.
\label{eq:difrad}
\end{equation}
We use $r_{dif}(E,E_g)$ to gain insight into which sources could contribute to the multi-TeV electron spectrum.  If a particle generated with even $E_g=100$~TeV is to be detected at $E=10$~TeV, it may propagate a maximal distance of $r_{dif}\sim\,$160 (230)~pc in slow (fast) diffusion models.  Note that the term $e^{-(r^2/r_{dif}^2)}/r_{dif}^{3}$ in Eq.~(\ref{eq:syrovatsky}) strongly favors particularly nearby sources.

%
\begin{figure}[t!]
\includegraphics[width=3.25in,clip=true]{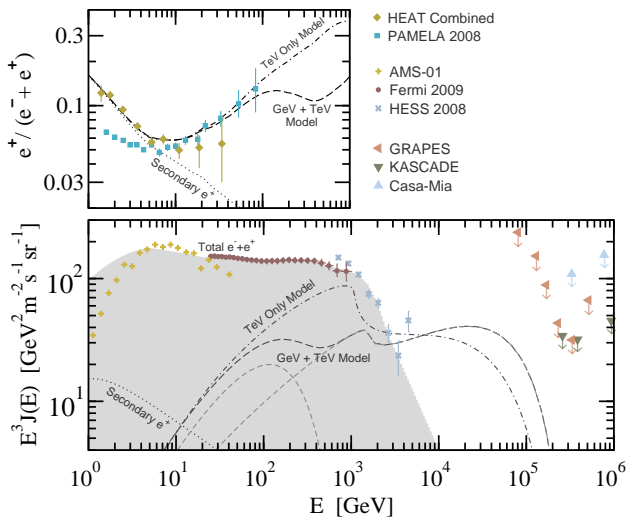}
\caption{Our modeled contributions to the $e^+$/$e^-$ spectra.
{\it Bottom:}
Shown are $e^- + e^+$ measurements from {\it Fermi}~\cite{Fermi:2009zk} and HESS~\cite{Aharonian:2008aaa,Aharonian:2009ah}; direct $e^-$ from AMS~\cite{Aguilar:2007yf}; our fit to these; and our $>10$~TeV limits from Fig.~\ref{limits}.  For Geminga, we show a model in which the TeV mode dominates and one with comparable GeV/TeV modes.
{\it Top:}
A comparison of the expected cosmic-ray positron fraction from the two models with data up to $\sim\,$100~GeV~\cite{Beatty:2004cy,Adriani:2008zr,Aguilar:2007yf}.
\label{electrons}}
\end{figure}
%

Closer inspection of the Syrovatskii solution reveals that more care is needed at these high energies.  For $E_g \sim E$, $E \simeq b_0\,t$ and $r_{dif}(t) \simeq 2 \sqrt{D(E)\,t}$, the quantity $v_{dif}(t)=r_{dif}(t)/t \simeq 2 \sqrt{D(E)/t}$ can exceed the speed of light.  This is the known problem of ``superluminal'' diffusion, which lies in the fact that the diffusion solution is not relativistic, similar to the Maxwell distribution yielding values in excess of $c$~\cite{Juttner(1911),Cubero(2007),Dunkel:2006kn}.  A phenomenological resolution was discussed in the context of ultrahigh-energy cosmic rays in Ref.~\cite{Aloisio:2008tx}, which uses a propagator based on the J\"{u}ttner particle distribution~\cite{Juttner(1911)} to explicitly limit fluxes to $v<c$, while preserving diffusive behavior at lower energies.  Now,
\begin{equation}
  n_J(r,t,E) = \frac{\theta(1-\xi)}{4 \pi (c t)^3}
         \frac{e^{-\alpha/\sqrt{1-\xi^2}}}{(1-\xi^2)^2} \frac{\alpha}{K_1(\alpha)}
            \frac{dN}{dE_g}\frac{dE_g}{dE} \,,
\label{eq:juttner}
\end{equation}
where $\xi(r,t) = r/ct$, $\theta$ is the step function, $K_1$ is the modified Bessel function, and $\alpha(E,t) = {c^2t^2}/({2\lambda(E,t)})$~\cite{Aloisio:2008tx}.

This formulation has the added benefit of removing spurious features appearing in the non-relativistic solution for source spectra harder than $E^{-2}$ and can easily be generalized to a continuously-emitting source at distance $r$ and time $t$ with a time-dependent particle injection rate $d\dot{N}/dE_g$, as $n_{\odot}(E)= \int_0^{t} dt^\prime\, \dot{n}(r,t^\prime,E)$.  Note that for sources such as pulsars, $r$ can vary with $t$~\cite{Yuksel:2008rf}.

\textit{Beyond the Positron Excess.}---
Using the J\"{u}ttner formalism, we can better estimate the extent to which a source may influence the spectrum at Earth.  For magnetic dipole braking, pulsar spin-down power evolves as $\propto (1+t/t_0)^{-(n+1)/(n-1)}$, with $n=3$ and a pulsar-dependent timescale, $t_0$.  Assuming that the $e^\pm$ output is simply proportional, the corresponding evolution will be ${\cal L}_{e^\pm}(t) = ({\cal E}_G/t_G)\, [1+(t_{G}-t)/t_{0}]^{-2}/\int^{t_G}dt' [1+(t_{G}-t')/t_{0}]^{-2}$.  We normalize to the $e^\pm$ luminosity of the source, ${\cal L}_{e^\pm}(t) = \int_{E_{\rm min}} E \, d\dot{N}/dE \,dE$ (with $E_{\rm min}=1\,$~GeV).

For Geminga, $\dot{E} \sim 10^{34.5}$~erg~s$^{-1}$ at present, and we use $t_0\sim 3\times 10^4$~yr (corresponding to a spin period at birth of $\sim\,$70~ms; see \cite{Atoyan:1995} for how this can vary).  We consider two distinct scenarios: a case from Ref.~\cite{Yuksel:2008rf} in which the TeV mode dominates and another with comparable energy in the dual GeV and TeV modes.  For the TeV spectra, we assume $E_{\rm max}=200\,$~TeV, and $\alpha=2$~\cite{Yuksel:2008rf}.  For the GeV spectrum, we use parameters consistent with Vela~X: $E_{\rm max}=150\,$~GeV, and $\alpha=1.8$.

%
\begin{table}[t!]
\caption{Parameters used for the two Geminga scenarios.}
\label{tab:params}
\begin{ruledtabular}
\begin{tabular}{lccccc}
Model & TeV mode          & GeV mode          & Distance	    & Age	   & Diffusion \\
      & (erg)             &	(erg)             & (pc)	        & (yr)	 &           \\ \hline
TeV   & $2\times 10^{48}$ & 0                 & 220           & $3\times10^5$ & slow \\
Dual  & $1.3\times 10^{48}$ & $1.3\times 10^{48}$ & 300$\rightarrow$200 & $2\times10^5$ & fast \\
\end{tabular}
\end{ruledtabular}
\end{table}
%

In Fig.~\ref{electrons}, we display the local flux of $e^- + e^+$, $J_\odot=(c/4\pi) \, n_\odot$, for these models, as described in Table~\ref{tab:params}.  These require $\sim 40$\% of the spin-down power be converted to high-energy $e^\pm$ pairs, within the range inferred from Vela~X~\cite{deJager:2008ni}.  We also examine the expected ratio of $e^+/(e^+ + e^-)$.  For the denominator, we directly use a fit to the measured data (neglecting the final HESS datum), as shown in Fig.~\ref{electrons}.  In the numerator, we include secondary fluxes~\cite{Moskalenko:1997gh} to match low-energy data.  As we see in the top panel, the two models can easily diverge at $\sim\,$100~GeV.  A more GeV-dominated case would drop more dramatically.

A few remarks are in order: (1) The underlying Galactic $e^-$ component is not known, but must be cut off at some point in order to not overshoot high-energy data (as seen in Fig.~\ref{electrons}).  This must make up the difference between any model and the full measured $e^- + e^+$ spectrum.  (2) Multiple nearby sources may contribute, although Geminga and Vela are the only bright gamma-ray pulsars within $\sim\,$300~pc.  (3) Low energies are more sensitive to the source's {\it initial} position, while at high energies the {\it present} position is most relevant (which can be directly measured).  (4) The emission properties (spectral index, cutoff, pair conversion efficiency, etc.) may evolve in time, while the highest-energy emission can reasonably be tied to what is seen today.  The observations of Vela do give hope of examining lost history, though.

\textit{Conclusions.}---
The flux level reached by our derived limits is encouraging for the prospects of upcoming generations of electron experiments, and already constrains the presence of a nearby, very-high-energy source.  Meanwhile, our handling of propagation in the presence of energy losses eliminates the appearance of spurious superluminal solutions.  This serves as a step towards a more accurate description of time-dependent $e^\pm$ propagation.  We have considered the nearby pulsar Geminga, detected in multi-TeV gamma rays, as the best motivated source for multi-TeV electrons reaching Earth today, although Vela~X may also contribute contingent on $e^\pm$ escape.

A nearby supernova remnant could result in electrons~\cite{Kobayashi:2003kp}, although age and distance play significant roles, with the recent history of acceleration being quite important due to the short cooling time at high energy.  Prominent SNRs include~\cite{Green:2009qf} the Vela SNR ($\sim\,$250~pc), Cyngus Loop ($\sim\,$440~pc), Monogem Ring ($\sim\,$300~pc), and Loop~I ($\sim\,$100~pc).  Due to their large angular extents, it is difficult to establish with ACTs whether these are active.  Searches via synchrotron radiation or wide-field gamma-ray instruments are better suited.  The only SNR potentially within 500~pc and detected in TeV gamma rays is Vela Junior, with an uncertain distance ranging from $\sim\,$200~pc (corresponding to an age of only $\sim\,$500~yr) to $\sim\,$1~kpc ($\sim\,$5000~yr)~\cite{Aharonian:2006dv}, while the ``Boomerang'' SNR/PWN at $\sim 800$~pc has been seen by Milagro~\cite{Abdo:2009ku}.

If features in the $e^- + e^+$ spectrum, such as the change in slope at $\sim 100$~GeV seen by {\it Fermi} or the drop at $\sim 1$~TeV measured by HESS, are due to a transition between different sources, classes, or $e^\pm$ populations, these should correspond to features in the positron fraction.  Measurements of the separate $e^-$/$e^+$ spectra will be vital in determining the Galactic component and in distinguishing SNR, dark matter~\cite{Hooper:2009cs}, and pulsar contributions.  For regions of pulsar domination, the positron fraction should saturate at $\sim 50$\%, while SNRs result entirely in primary electrons (unless secondary acceleration occurs~\cite{Blasi:2009bd}).  While PAMELA can reach 300~GeV for $e^+$~\cite{Boezio:2009}, AMS-02 will go to $\sim 1$~TeV~\cite{Beischer:2009}, and ACT measurements using the varying position of the moon shadow for $e^+$/$e^-$ may reach several TeV~\cite{Colin:2009fs}.  These combined observations give hope for connecting very-high-energy emission to lower energies to build a complete picture of the spectra of cosmic-ray electron and positrons.

%
We thank Jim Beatty, Ty DeYoung, Tom Gaisser, Jamie Holder, Amir Javaid, Brian Lacki, Dave Seckel, Todor Stanev, Todd Thompson, and especially John Beacom for discussions and comments.
MDK is supported by an OSU Presidential Fellowship and NSF CAREER grant PHY-0547102 (to JFB)
and HY by DOE grant DE-FG02-91ER40626.



\vspace*{-0.6cm}
\begin{thebibliography}{99}
\vspace*{-0.6cm}

\bibitem{Adriani:2008zr}
  O.~Adriani {\it et al.},
  Nature, {\bf 458}, 607 (2009).

\bibitem{Fermi:2009zk}
  A.~A.~Abdo {\it et al.}
  Phys.\ Rev.\ Lett.\  {\bf 102}, 181101 (2009).

\bibitem{Aharonian:2008aaa}
  F.~Aharonian {\it et al.},
  Phys.\ Rev.\ Lett.\  {\bf 101}, 261104 (2008).

\bibitem{Aharonian:2009ah}
  F.~Aharonian {\it et al.},
  arXiv:0905.0105.

\bibitem{Yuksel:2008rf}
  H.~Yuksel, M.~D.~Kistler and T.~Stanev,
  Phys.\ Rev.\ Lett.\  {\bf 103}, 051101 (2009).

\bibitem{Hooper:2008kg}
  D.~Hooper, P.~Blasi and P.~D.~Serpico,
  JCAP {\bf 0901}, 025 (2009).

\bibitem{Profumo:2008ms}
  S.~Profumo,
  arXiv:0812.4457.

\bibitem{Blasi:2009bd}
  P.~Blasi and P.~D.~Serpico,
  Phys.\ Rev.\ Lett.\  {\bf 103}, 081103 (2009);
%
  P.~Mertsch and S.~Sarkar,
  Phys.\ Rev.\ Lett.\  {\bf 103}, 081104 (2009).

\bibitem{Chang:2008zzr}
  J.~Chang {\it et al.},
  Nature {\bf 456}, 362 (2008).

\bibitem{ArkaniHamed:2008qn}
  N.~Arkani-Hamed, D.~P.~Finkbeiner, T.~Slatyer and N.~Weiner,
  Phys.\ Rev.\  D {\bf 79}, 015014 (2009);
%
  M.~Pospelov and A.~Ritz,
  Phys.\ Lett.\  B {\bf 671}, 391 (2009);
%
  M.~Cirelli, M.~Kadastik, M.~Raidal and A.~Strumia,
  Nucl.\ Phys.\  B {\bf 813}, 1 (2009);
%
  Y.~Nomura and J.~Thaler,
  Phys.\ Rev.\  D {\bf 79}, 075008 (2009);
%
  I.~Cholis, G.~Dobler, D.~P.~Finkbeiner, L.~Goodenough and N.~Weiner,
  arXiv:0811.3641.
%
  D.~Feldman, Z.~Liu and P.~Nath,
  Phys.\ Rev.\  D {\bf 79}, 063509 (2009);
%
  V.~Barger, W.~Y.~Keung, D.~Marfatia and G.~Shaughnessy,
  Phys.\ Lett.\  B {\bf 672}, 141 (2009);
%
  M.~Ibe, Y.~Nakayama, H.~Murayama and T.~T.~Yanagida,
  JHEP {\bf 0904}, 087 (2009).



\bibitem{Moskalenko:1997gh}
  I.~V.~Moskalenko and A.~W.~Strong,
  Astrophys.\ J.\  {\bf 493}, 694 (1998).

\bibitem{Hofmann:2006wf}
  W.~Hofmann,
  astro-ph/0603076.

\bibitem{Aharonian:2001ft}
  F.~A.~Aharonian {\it et al.},
  Astropart.\ Phys.\  {\bf 17}, 459 (2002).

\bibitem{Chantell:1997gs}
  M.~C.~Chantell {\it et al.},
  Phys.\ Rev.\ Lett.\  {\bf 79}, 1805 (1997).

\bibitem{Gupta}
  S.~Gupta {\it et al.},
  Proc.\ 31st Intl.\ Cosmic Ray Conf., (2009).

\bibitem{Schatz:2003aw}
  G.~Schatz {\it et al.},
  Proc.\ 28th Intl.\ Cosmic Ray Conf., Tsukuba, {\bf 4}, 2293 (2003).


\bibitem{Aharonian:2008zz}
  F.~Aharonian, J.~Buckley, T.~Kifune and G.~Sinnis,
  Rept.\ Prog.\ Phys.\  {\bf 71}, 096901 (2008).

\bibitem{Kistler:2009xf}
  M.~D.~Kistler and J.~M.~Siegal-Gaskins,
  arXiv:0909.0519.

\bibitem{Hillas:2006ms}
  A.~M.~Hillas,
  arXiv:astro-ph/0607109.


\bibitem{Abdo:2009ku}
  A.~A.~Abdo {\it et al.},
  Astrophys.\ J.\  {\bf 700}, L127 (2009).

\bibitem{Aharonian:2006zb}
  F.~Aharonian {\it et al.},
  Astron.\ Astrophys.\  {\bf 460}, 365 (2006).

\bibitem{Aharonian:2006xx}
  F.~Aharonian {\it et al.},
  Astron.\ Astrophys.\  {\bf 448}, L43 (2006).

\bibitem{deJager:2008ni}
  O.~C.~de Jager, P.~O.~Slane and S.~LaMassa,
  Astrophys.\ J.\  {\bf 689}, L125 (2008).

\bibitem{Lemoine}
  M.~Lemoine-Goumard,
  ``Fermi-LAT Observations of the Vela X Region,''
  talk at the 2009 Fermi Symposium,
  http://fermi.gsfc.nasa.gov/science/symposium/2009/.

\bibitem{deJager(2007)}
  O.~C.~de Jager,
  Astrophys.\ J.\  {\bf 658}, 1177 (2007).

\bibitem{Ginzburg(1964)}
  V.~L.~Ginzburg and S.~I.~Syrovatskii,
  {\it The Origin of Cosmic Rays}, (Macmillan, New York, 1964).

\bibitem{Strong:2007nh}
  A.~W.~Strong, I.~V.~Moskalenko and V.~S.~Ptuskin,
  Ann.\ Rev.\ Nucl.\ Part.\ Sci.\  {\bf 57}, 285 (2007).

\bibitem{Aharonian(1995)}
  F.~Aharonian, A.~M.~Atoyan and H.~J.~Volk,
  Astron.\ Astrophys.\  {\bf 294}, L41 (1995).

\bibitem{DiBernardo:2009ku}
  G.~Di Bernardo {\it et al.},
  arXiv:0909.4548.

\bibitem{Juttner(1911)}
  F.~J{\"u}ttner,
  Annalen Phys. {\bf 339}, 856 (1911).

\bibitem{Cubero(2007)}
  D.~Cubero {\it et al.},
  Phys.\ Rev.\ Lett.\ {\bf 99}, 170601 (2007).

\bibitem{Dunkel:2006kn}
  J.~Dunkel {\it et al.},
  Phys.\ Rev.\  D {\bf 75}, 043001 (2007).

\bibitem{Aloisio:2008tx}
  R.~Aloisio, V.~Berezinsky and A.~Gazizov,
  Astrophys.\ J.\  {\bf 693}, 1275 (2009).

\bibitem{Atoyan:1995}
  A.~M.~Atoyan, F.~A.~Aharonian and H.~J.~Volk,
  Phys.\ Rev.\  D {\bf 52}, 3265 (1995).

\bibitem{Kobayashi:2003kp}
  T.~Kobayashi {\it et al.},
  Astrophys.\ J.\  {\bf 601}, 340 (2004).

\bibitem{Green:2009qf}
  D.~A.~Green,
  arXiv:0905.3699.

\bibitem{Aharonian:2006dv}
  F.~Aharonian {\it et al.},
  Astrophys.\ J.\  {\bf 661}, 236 (2007).

\bibitem{Hooper:2009cs}
  D.~Hooper and K.~M.~Zurek,
  arXiv:0909.4163;
%
  I.~Cholis and N.~Weiner,
  arXiv:0911.4954.

\bibitem{Boezio:2009}
  M.~Boezio {\it et al.},
  New J.\ Phys.\ {\bf 11}, 105023 (2009).

\bibitem{Beischer:2009}
  B.~Beischer {\it et al.},
  New J.\ Phys.\ {\bf 11}, 105021 (2009).

\bibitem{Colin:2009fs}
  P.~Colin {\it et al.} [MAGIC Collaboration],
  arXiv:0907.1026.
  
\bibitem{Beatty:2004cy}
  J.~J.~Beatty {\it et al.},
  Phys.\ Rev.\ Lett.\  {\bf 93}, 241102 (2004).

\bibitem{Aguilar:2007yf}
  M.~Aguilar {\it et al.},
  Phys.\ Lett.\  B {\bf 646}, 145 (2007).



\end{thebibliography}
\end{document}